\documentstyle[preprint,eqsecnum,aps]{revtex}
\begin{document}
\draft
\preprint{}
\title{``Faster Than Light" Photons in Dilaton Black Hole Spacetimes}
\author{
H. T. Cho \footnote[1]{htcho@mail.tku.edu.tw }}
\address{Tamkang University, Department of Physics,\\
Tamsui, Taipei, TAIWAN R.O.C.}
\maketitle
\begin{abstract}

We investigate the phenomenon of ``faster than light" photons in
a family of dilaton black hole spacetimes. For radially directed
photons, we find that their light-cone condition is modified 
even though the spacetimes are spherically symmetric. They also
satisfy the ``horizon theorem" and the ``polarization sum rule" 
of Shore. For orbital photons, the dilatonic effect on the 
modification of the light-cone condition can become 
more dominant than the electromagnetic and the gravitational ones
as the orbit gets closer to the event horizon 
in the extremal or near-extremal cases.

\end{abstract}

\section{Introduction}

In 1980, Drummond and Hathrell \cite{DH} discovered that photons may 
propagate faster than the ``speed of light" $c$ in curved spacetimes
if QED one-loop quantum effects are taken into account. For instance,
they found that in a Schwarzschild black hole spacetime, an orbital
photon is superluminal in one polarization and subluminal in the other. 
This in fact gives a gravitational analogue to the phenomenon of 
electromagnetic birefringence \cite{SLA}. Due to the one-loop vacuum 
polarization, 
the photon exists part of the time as a virtual $e^{+}e^{-}$ pair. This
introduces a size on the photon of the order of the Compton wavelength
of the electron $\lambda_{c}=1/m$, where $m$ is the mass of the
electron (Here we use the notation $\hbar=c=\epsilon_{0}=1$.) After
acquiring a size, the motion of the photon could thus be altered by the
tidal effects of the spacetime curvature. As discussed in Ref. 1, however,
this change of speed does not necessarily imply any violation of causality.

In the past few years, the study of this phenomenon has been 
extended by Daniels and Shore to Reissner-Nordstr\"om (RN) \cite{DS1}
and Kerr \cite{DS2}
black hole spacetimes. Two general features, called the ``horizon
theorem" and the ``polarization sum rule", emerge from these
considerations \cite{GMS}. First, the horizon theorem states that 
the velocity of radial photons remains equal 
to $c$ at the event horizon. In fact it has been found that 
in spherically symmetric spacetimes like the Schwarzschild and RN 
ones, the velocity
of radial photons does not change at all. Second, the polarization sum
rule states that the polarization
averaged velocity shift is proportional to the matter energy-momentum
tensor. Therefore velocity shifts of the
two polarizations are equal and opposite in Ricci flat spacetimes.

In this work we would like to study this phenomenon in dilaton
black hole spacetimes. Dilaton gravity arises from low-energy effective
string actions. The corresponding black hole solutions \cite{GM,GHS}
exhibit quite different causal structures from the usual 
RN solutions. For example, the inner horizon is
a spacelike surface of singularity in the dilaton case versus a regular
surface in the usual one. Therefore it would be interesting to see if
there is any peculiarity in the propagation of photons in these rather 
different black hole spacetimes. 

In the next section we consider the effective action of dilaton gravity 
which includes one-loop quantum effects from matter. Using the geometric
optics approximation, we derive the equation of photon propagation. In 
Section III, the propagations of photons in different dilaton black hole 
spacetimes, parametrized by the dilaton coupling constant $\hat{a}$, 
are studied. The light-cone condition of the specific cases of radial 
and orbital photons are considered in more details in Sections IV and
V, respectively. Conclusions
and discussions are presented in Section VI.

\section{Photon propagation in dilaton gravity}

We shall be using the action for dilaton gravity,
\begin{equation}
S=\int d^{4}x\sqrt{-g}[R-2(\nabla\phi)^{2}-e^{-2\hat{a}\phi}F^{2}],
\label{action}
\end{equation}
where $\phi$ is the dilaton field and $\hat{a}$ is the dilaton
coupling. The one-loop quantum effects from matter is
summarized in the effective action \cite{DS1}
\begin{eqnarray}
S_{1}&=&\frac{1}{m^2}\int d^{4}x\sqrt{-g}[aRF_{\mu\nu}F^{\mu\nu}+
bR_{\mu\nu}F^{\mu\sigma}{F^{\nu}}_{\sigma}
+cR_{\mu\nu\sigma\tau}F^{\mu\nu}F^{\sigma\tau}+
d(\nabla_{\mu}F^{\mu\nu})(\nabla_{\sigma}{F^{\sigma}}_{\nu})]
\nonumber\\
&&+\frac{1}{m^4}\int d^{4}x\sqrt{-g}[z(F_{\mu\nu}F^{\mu\nu})^{2}+
yF_{\mu\nu}F_{\sigma\tau}F^{\mu\sigma}F^{\nu\tau}],
\end{eqnarray}
where $a, b, c, d, z$ and $y$ are constants. For QED corrections,
\begin{equation}
\begin{array}{lll}
a=\frac{\alpha}{36\pi} & b=-\frac{13\alpha}{90\pi} & 
c=\frac{\alpha}{90\pi} \\
d=\frac{2\alpha}{15\pi} & z=-\frac{\alpha^2}{9} & 
y=\frac{14\alpha^{2}}{45}
\end{array}
\end{equation}
where $\alpha$ is the fine structure constant and $m$ is the 
electron mass. Here we are effectively making a local expansion 
in powers of $R/m^2$ and $\alpha F^{2}/m^{4}$ where $R$ and $F$ 
are generic curvature and field strength, and are retaining only 
the leading order terms.   

We derive the equation of motion for the electromagnetic field by 
taking the variation
\begin{eqnarray}
&&\frac{\delta(S+S_{1})}{\delta A_{\nu}}=0
\nonumber\\
&\Longrightarrow&\nabla_{\mu}(e^{-2\hat{a}\phi}F^{\mu\nu})
\nonumber\\
&&-\frac{1}{m^{2}}a\nabla_{\mu}(RF^{\mu\nu})
-\frac{1}{2m^2}b\nabla_{\sigma}(R^{\mu\sigma}{F_{\mu}}^{\nu}-
R^{\mu\nu}{F_{\mu}}^{\sigma})
\nonumber\\
&&-\frac{1}{m^2}c\nabla_{\mu}({R^{\mu\nu}}_{\sigma\tau}F^{\sigma\tau})
+\frac{1}{2m^2}d(\Box\nabla_{\sigma}F^{\sigma\nu}- 
\nabla^{\mu}\nabla^{\nu}\nabla_{\sigma}{F^{\sigma}}_{\mu})
\nonumber\\
&&-\frac{2}{m^4}z(F^{\sigma\tau}F_{\sigma\tau}
\nabla_{\mu}F^{\mu\nu}+2F^{\mu\nu}F_{\sigma\tau}
\nabla_{\mu}F^{\sigma\tau})
\nonumber\\
&&-\frac{2}{m^4}y(F^{\nu\tau}F_{\sigma\tau}\nabla_{\mu}
F^{\mu\sigma}+F^{\mu\sigma}F_{\sigma\tau}\nabla_{\mu}
F^{\nu\tau}+F^{\mu\sigma}F^{\nu\tau}\nabla_{\mu}
F_{\sigma\tau})=0.
\label{vary}
\end{eqnarray}
Next we write 
\begin{equation}
F_{\mu\nu}=\bar{F}_{\mu\nu}+\hat{f}_{\mu\nu},
\label{photon}
\end{equation}
where $\bar{F}_{\mu\nu}$ is the background electromagnetic 
field and $\hat{f}_{\mu\nu}$ is the photon field. To obtain
the equation of motion for the photon, we put Eq.(\ref{photon}) into 
Eq.(\ref{vary}) and take the part linear in $\hat{f}_{\mu\nu}$. In 
addition, we assume that typical variations of the background
electromagnetic, gravitational and dilatonic fields, 
characterized by the scale $L$, are much smaller than that of
the photon field, that is,
\begin{equation}
L\gg\lambda
\end{equation}
where $\lambda$ is the photon wavelength. Then derivatives like
$\nabla\bar{F}$ and $\nabla R$ can be neglected as compared to
$\nabla\hat{f}$. Hence, from Eq.(\ref{vary}) we have,
\begin{eqnarray}
&&\nabla_{\mu}\hat{f}^{\mu\nu}
-\frac{1}{m^{2}}ae^{2\hat{a}\phi}R(\nabla_{\mu}\hat{f}^{\mu\nu})
-\frac{1}{2m^2}be^{2\hat{a}\phi}(R^{\mu\sigma}\nabla_{\sigma}
{\hat{f}_{\mu}}^{\ \nu}-R^{\mu\nu}\nabla_{\sigma}
{\hat{f}_{\mu}}^{\ \sigma})
\nonumber\\
&&-\frac{1}{m^2}ce^{2\hat{a}\phi}{R^{\mu\nu}}_{\sigma\tau}
(\nabla_{\mu}\hat{f}^{\sigma\tau})
+\frac{1}{2m^2}de^{2\hat{a}\phi}(\Box\nabla_{\sigma}\hat{f}^{\sigma\nu}- 
\nabla^{\mu}\nabla^{\nu}\nabla_{\sigma}{\hat{f}^{\sigma}}_{\ \mu})
\nonumber\\
&&-\frac{2}{m^4}ze^{2\hat{a}\phi}(\bar{F}^{\sigma\tau}\bar{F}_{\sigma\tau}
\nabla_{\mu}\hat{f}^{\mu\nu}+2\bar{F}^{\mu\nu}\bar{F}_{\sigma\tau}
\nabla_{\mu}\hat{f}^{\sigma\tau})
\nonumber\\
&&-\frac{2}{m^4}ye^{2\hat{a}\phi}
(\bar{F}^{\nu\tau}\bar{F}_{\sigma\tau}\nabla_{\mu}
\hat{f}^{\mu\sigma}+\bar{F}^{\mu\sigma}\bar{F}_{\sigma\tau}\nabla_{\mu}
\hat{f}^{\nu\tau}+\bar{F}^{\mu\sigma}\bar{F}^{\nu\tau}\nabla_{\mu}
\hat{f}_{\sigma\tau})=0.
\end{eqnarray}
Without quantum corrections, $\nabla_{\mu}\hat{f}^{\mu\nu}=0$. 
Thus $\nabla_{\mu}\hat{f}^{\mu\nu}$ is at least of first order 
(order of $\alpha$ in QED) in perturbation. Since we are 
considering only first order effects, we should consistently
neglect terms in the above equation with 
$\nabla_{\mu}\hat{f}^{\mu\nu}$ which are of second order or higher,
giving
\begin{eqnarray}
&&\nabla_{\mu}\hat{f}^{\mu\nu}
-\frac{1}{2m^2}be^{2\hat{a}\phi}(R^{\mu\sigma}\nabla_{\sigma}
{\hat{f}_{\mu}}^{\ \nu})
-\frac{1}{m^2}ce^{2\hat{a}\phi}{R^{\mu\nu}}_{\sigma\tau}
(\nabla_{\mu}\hat{f}^{\sigma\tau})
\nonumber\\
&&-\frac{4}{m^4}ze^{2\hat{a}\phi}\bar{F}^{\mu\nu}\bar{F}_{\sigma\tau}
\nabla_{\mu}\hat{f}^{\sigma\tau}
-\frac{2}{m^4}ye^{2\hat{a}\phi}
(\bar{F}^{\mu\sigma}\bar{F}_{\sigma\tau}\nabla_{\mu}
\hat{f}^{\nu\tau}+\bar{F}^{\mu\sigma}\bar{F}^{\nu\tau}\nabla_{\mu}
\hat{f}_{\sigma\tau})=0.
\end{eqnarray}
Actually, to neglect terms containing $\nabla_{\mu}\hat{f}^{\mu\nu}$,
we need to have a bound on this derivative. A sufficient condition
is that,
\begin{equation}
\lambda\gg\lambda_{c},
\end{equation}
where $\lambda_{c}$ is the electron Compton wavelength.

To study the propagation of the photon, we use the geometric
optics approximation \cite{MTW}, in which one writes
\begin{equation}
\hat{f}_{\mu\nu}=f_{\mu\nu}e^{i\theta},
\end{equation}
where $\theta$ is a rapidly varying phase and $f_{\mu\nu}$ 
a slowly varying amplitude. The momentum of the photon is
given by $k_{\mu}=\nabla_{\mu}\theta$ and the amplitude can 
be written as 
\begin{equation}
f_{\mu\nu}=k_{\mu}a_{\nu}-k_{\nu}a_{\mu},
\end{equation}
where $a_{\mu}$ is the polarization vector satisfying the condition,
\begin{equation}
k_{\mu}a^{\mu}=0.
\end{equation}
In this geometric optics approximation, the equation of motion becomes
\begin{eqnarray}
&&k_{\mu}f^{\mu\nu}
-\frac{1}{2m^2}be^{2\hat{a}\phi}(R^{\mu\sigma}k_{\sigma}{f_{\mu}}^{\ \nu})
-\frac{1}{m^2}ce^{2\hat{a}\phi}{R^{\mu\nu}}_{\sigma\tau}
(k_{\mu}f^{\sigma\tau})
\nonumber\\
&&-\frac{4}{m^4}ze^{2\hat{a}\phi}\bar{F}^{\mu\nu}\bar{F}_{\sigma\tau}
k_{\mu}f^{\sigma\tau}
-\frac{2}{m^4}ye^{2\hat{a}\phi}
(\bar{F}^{\mu\sigma}\bar{F}_{\sigma\tau}k_{\mu}f^{\nu\tau}
+\bar{F}^{\mu\sigma}\bar{F}^{\nu\tau}k_{\mu}f_{\sigma\tau})=0
\nonumber\\
&\Longrightarrow&k^{2}a^{\nu}
-\frac{1}{2m^2}be^{2\hat{a}\phi}R^{\mu\sigma}k_{\sigma}
(k_{\mu}a^{\nu}-k^{\nu}a_{\mu})
-\frac{2}{m^2}ce^{2\hat{a}\phi}{R^{\mu\nu}}_{\sigma\tau}
k_{\mu}k^{\sigma}a^{\tau}
\nonumber\\
&&-\frac{8}{m^4}ze^{2\hat{a}\phi}\bar{F}^{\mu\nu}\bar{F}_{\sigma\tau}
k_{\mu}k^{\sigma}a^{\tau}
-\frac{2}{m^4}ye^{2\hat{a}\phi}
(\bar{F}^{\mu\sigma}\bar{F}_{\sigma\tau}k_{\mu}
(k^{\nu}a^{\tau}-k^{\tau}a^{\nu})
-\bar{F}^{\mu\sigma}\bar{F}^{\nu\tau}k_{\mu}
k_{\tau}a_{\sigma})=0.
\nonumber\\
&&
\label{motiont}
\end{eqnarray}
In the next section, we shall study the propagation of photons
in dilaton black hole spacetimes using these equations.

\section{Dilaton black hole spacetimes}

One family of electrically charged dilaton black holes \cite{GM,GHS}
can be obtained from the action in Eq.(\ref{action}) with the line element
\begin{equation}
ds^{2}=-\lambda^{2}dt^{2}+\lambda^{-2}dr^{2}
+R^{2}(d\theta^{2}+{\rm sin}^{2}\theta d\phi^{2}),
\end{equation}
where 
\begin{eqnarray}
\lambda&=&\left(1-\frac{r_{+}}{r}\right)^{1/2}
\left(1-\frac{r_{-}}{r}\right)^{(1-\hat{a}^{2})/2(1+\hat{a}^{2})},
\\
R&=&r\left(1-\frac{r_{-}}{r}\right)^{\hat{a}^{2}/(1+\hat{a}^{2})},
\end{eqnarray}
with $r=r_{+}$ and $r=r_{-}$ corresponding 
to the outer and inner horizons, respectively.
$r_{+}$ and $r_{-}$ are related
to the mass $M$ and charge $Q$ of the black hole by
\begin{eqnarray}
2GM&=&r_{+}+\left(\frac{1-\hat{a}^{2}}{1+\hat{a}^{2}}\right)r_{-},
\\
\frac{GQ^{2}}{4\pi}&=&\frac{r_{+}r_{-}}{1+\hat{a}^{2}},
\end{eqnarray}
where $G$ is the Newton's constant.
Moreover, the electric and the dilaton fields are given by
\begin{eqnarray}
\bar{F}_{01}&=&\frac{Q}{4\pi r^{2}},
\\
e^{2\hat{a}\phi}
&=&\left(1-\frac{r_{-}}{r}\right)^{2\hat{a}^{2}/(1+\hat{a}^{2})}.
\end{eqnarray}

To introduce a local set of orthonormal frames, we use the 
vierbein fields ${e_{\mu}}^{a}$ defined by
\begin{equation}
g_{\mu\nu}=\eta_{ab}{e_{\mu}}^{a}{e_{\nu}}^{b},
\end{equation}
where $\eta_{ab}$ is the Minkowski metric, and
\begin{equation}
{e_{\mu}}^{a}={\rm diag}(\lambda,1/\lambda,R,R\,{\rm sin}\theta),
\end{equation}
with the inverse
\begin{equation}
{e_{a}}^{\mu}={\rm diag}(1/\lambda,\lambda,1/R,1/R\,{\rm sin}\theta).
\end{equation}
The vierbein components of the Riemann tensor $R_{abcd}$ are
\begin{eqnarray}
&&R_{0101}=\frac{d}{dr}\left(\lambda
\frac{d\lambda}{dr}\right)\equiv A,
\\
&&R_{0202}=R_{0303}=\frac{\lambda}{R}\left(\frac{dR}{dr}\right)
\left(\frac{d\lambda}{dr}\right)\equiv B,
\\
&&R_{1212}=R_{1313}=-\frac{\lambda^{2}}{R}
\left(\frac{d^{2}R}{dr^{2}}\right)-\frac{\lambda}{R}
\left(\frac{dR}{dr}\right)\left(\frac{d\lambda}{dr}\right)\equiv C,
\\
&&R_{2323}=\frac{1}{R^{2}}-\frac{\lambda^{2}}{R^{2}}
\left(\frac{dR}{dr}\right)^{2}\equiv D.
\nonumber
\end{eqnarray}
Using the notation $U^{01}_{ab}\equiv\delta^{0}_{a}
\delta^{1}_{b}-\delta^{1}_{a}\delta^{0}_{b}$, etc., the Riemann
tensor can be expressed more compactly as 
\begin{equation}
R_{abcd}=AU^{01}_{ab}U^{01}_{cd}+
B(U^{02}_{ab}U^{02}_{cd}+U^{03}_{ab}U^{03}_{cd})+
C(U^{12}_{ab}U^{12}_{cd}+U^{13}_{ab}U^{13}_{cd})+
DU^{23}_{ab}U^{23}_{cd},
\end{equation}
and the vierbein components of the background electric field is
\begin{equation}
\bar{F}_{ab}=\frac{Q}{4\pi r^{2}}U^{01}_{ab}.
\end{equation}
In this local set of orthonormal frames, the equation of
motion for photons in Eq.(\ref{motiont}) can be written as 
\begin{eqnarray}
&&k^{2}a^{b}
-\frac{1}{2m^2}be^{2\hat{a}\phi}R^{ac}k_{c}(k_{a}a^{b}-k^{b}a_{a})
-\frac{2}{m^2}ce^{2\hat{a}\phi}{R^{ab}}_{cd}k_{a}k^{c}a^{d}
\nonumber\\
&&-\frac{8}{m^4}ze^{2\hat{a}\phi}\bar{F}^{ab}\bar{F}_{cd}k_{a}k^{c}a^{d}
-\frac{2}{m^4}ye^{2\hat{a}\phi}
(\bar{F}^{ac}\bar{F}_{cd}k_{a}(k^{b}a^{d}-k^{d}a^{b})
-\bar{F}^{ac}\bar{F}^{bd}k_{a}k_{d}a_{c})=0.
\label{motionv}
\end{eqnarray}
The independent components of the polarization vector can be
projected out using the vectors,
\begin{eqnarray}
l_{b}&=&k^{a}U^{01}_{ab},
\\
m_{b}&=&k^{a}U^{02}_{ab},
\\
n_{b}&=&k^{a}U^{03}_{ab}.
\end{eqnarray}
since $l^{a}$, $m^{a}$ and $n^{a}$ are independent and orthogonal to $k^{a}$. We also
introduce the dependent vectors
\begin{eqnarray}
p_{b}&=&k^{a}U^{12}_{ab}=\frac{1}{k^{0}}(k^{1}m_{b}-k^{2}l_{b})
\\
q_{b}&=&k^{a}U^{13}_{ab}=\frac{1}{k^{0}}(k^{1}n_{b}-k^{3}l_{b})
\\
r_{b}&=&k^{a}U^{23}_{ab}=\frac{1}{k^{0}}(k^{2}n_{b}-k^{3}m_{b})
\end{eqnarray}
Contracting the equations of motion in Eq.(\ref{motionv}) with $l^{a}$, 
$m^{a}$ and $n^{a}$, respectively, we obtain equations for each independent
components of the polarization vector $a^{a}$,
\begin{eqnarray}
&&k^{2}(a\cdot v)-\frac{1}{2m^{2}}be^{2\hat{a}\phi}
[Al^{2}+B(m^{2}+n^{2})+C(p^{2}+q^{2})+Dr^{2}](a\cdot v)
\nonumber\\
&&\ -\frac{2}{m^{2}}ce^{2\hat{a}\phi}[A(l\cdot v)(a\cdot l)
+B(m\cdot v)(a\cdot m)+B(n\cdot v)(a\cdot n)
\nonumber\\
&&\ \ \ \ \ \ +C(p\cdot v)(a\cdot p)+C(q\cdot v)(a\cdot q)
+D(r\cdot v)(a\cdot r)]
\nonumber\\
&&\ -\frac{8}{m^{4}}ze^{2\hat{a}\phi}\left(\frac{Q}{4\pi r^{2}}\right)^{2}
(l\cdot v)(a\cdot l)
-\frac{2}{m^{4}}ye^{2\hat{a}\phi}\left(\frac{Q}{4\pi r^{2}}\right)^{2}
[l^{2}(a\cdot v)+(l\cdot v)(a\cdot l)]=0,
\label{polar}
\end{eqnarray}
where $v^{a}=l^{a}$, $m^{a}$ or $n^{a}$.
To show the peculiarities of the propagation of photons in these 
situations we shall concentrate on the cases of radially directed
and orbital photons in the next two sections.

\section{Radial photons}

For radially directed photons,
\begin{equation}
k^{2}=k^{3}=0,
\end{equation}
then
\begin{eqnarray}
l^{a}&=&(k^{1},k^{0},0,0),
\\
m^{a}&=&(0,0,k^{0},0),
\\
n^{a}&=&(0,0,0,k^{0}).
\end{eqnarray}
Thus $(a\cdot l)$ corresponds to the unphysical polarization. On the 
other hand, the equations for the physical polarizations 
$(a\cdot m)$ and $(a\cdot n)$ in Eq.(\ref{polar}) are the same,
\begin{eqnarray}
&&\left\{k^{2}-\frac{1}{2m^{2}}be^{2\hat{a}\phi}
\left[Al^{2}+B(m^{2}+n^{2})+C(p^{2}+q^{2})\right]\right.
\nonumber\\
&&\ -\frac{2}{m^{2}}ce^{2\hat{a}\phi}\left[Bn^{2}
+C\left(\frac{k^{1}}{k^{0}}\right)^{2}n^{2}\right]
\left. -\frac{2}{m^{4}}ye^{2\hat{a}\phi}\left(\frac{Q}{4\pi r^{2}}\right)^{2}
l^{2}\right\}(a\cdot v)=0,
\end{eqnarray}
where $v^{a}=m^{a}$ or $n^{a}$. 

To have non-trivial solutions, we require
\begin{eqnarray}
&&k^{2}-\frac{1}{2m^{2}}be^{2\hat{a}\phi}
[Al^{2}+B(m^{2}+n^{2})+C(p^{2}+q^{2})]
\nonumber\\
&&\ -\frac{2}{m^{2}}ce^{2\hat{a}\phi}\left[Bn^{2}
+C\left(\frac{k^{1}}{k^{0}}\right)^{2}n^{2}\right]
-\frac{2}{m^{4}}ye^{2\hat{a}\phi}\left(\frac{Q}{4\pi r^{2}}\right)^{2}
l^{2}=0.
\end{eqnarray}
To first order in perturbation, this gives
\begin{eqnarray}
\left\vert\frac{k^{0}}{k^{1}}\right\vert
&=&1-\frac{1}{2m^{2}}(b+2c)e^{2\hat{a}\phi}(B+C)
\nonumber\\
&=&1-\frac{1}{2m^{2}}(b+2c)\frac{\hat{a}^{2}}
{(1+\hat{a}^{2})^{2}}\left(\frac{r_{-}^{2}}{r^{4}}\right)
\left(1-\frac{r^{+}}{r}\right)\left(1-\frac{r^{-}}{r}\right)^{-1}.
\label{speed}
\end{eqnarray}
For $\hat{a}^{2}=0$,
\begin{equation}
\left\vert\frac{k^{0}}{k^{1}}\right\vert=1,
\end{equation}
which is the case for Schwarzschild or RN black holes
as well as other spherically symmetric spacetimes \cite{DH,DS1}, 
where the light-cone condition for radial photons is not 
modified.
For $\hat{a}^{2}\neq0$, 
the light-cone condition
is modified in this dilatonic case 
even though the spacetime
is still spherically symmetric. 
For example, for QED quantum corrections,
\begin{eqnarray}
b+2c&=&-\frac{13\alpha}{90\pi}+\frac{2\alpha}{90\pi}
\nonumber\\
&=&-\frac{11\alpha}{90\pi}<0,
\end{eqnarray}
the radial photons in fact propagate superluminally when
$r>r_{+},r_{-}$.

At the event horizon, $r=r_{+}$,
\begin{equation}
\left\vert\frac{k^{0}}{k^{1}}\right\vert=1,
\end{equation}
which is in accordance with the ``horizon theorem" proven
by Shore \cite{GMS}, that the light-cone condition for radial
photons becomes $k^{2}=0$ at the event horizon.
From Eq.(\ref{speed}), we also see that at the inner horizon, $r=r_{-}$,
the velocity shift diverges. This reflects the fact that the
inner horizon for dilaton black holes is actually a singular
surface.

We can also consider the other theorem of Shore \cite{GMS}, the 
polarization sum rule. For radial photons, to first order in
perturbation,
\begin{eqnarray}
R_{ab}k^{a}k^{b}&=&Al^{2}+B(m^{2}+n^{2})+C(p^{2}+q^{2})
\nonumber\\
&=&2(B+C)(k^{1})^{2}.
\label{ricci}
\end{eqnarray}
Using Eqs.(\ref{ricci}) and (\ref{speed}), 
we have for the two polarizations, 
\begin{eqnarray}
\sum_{{\rm pol}}k^{2}&=&\sum_{{\rm pol}}[-(k^{0})^{2}+(k^{1})^{2}]
\nonumber\\
&=&\frac{1}{m^{2}}(b+2c)e^{2\hat{a}\phi}[2(B+C)(k^{1})^{2})]
\nonumber\\
&=&\frac{1}{m^{2}}(b+2c)e^{2\hat{a}\phi}R_{ab}k^{a}k^{b},
\end{eqnarray}
which is just the ``polarization sum rule" considered by Shore, 
modified to the dilaton case.

\section{Orbital photons}
For orbital photons, take
\begin{equation}
k^{1}=k^{2}=0\ \ \ {\rm and}\ \ \ \theta=\frac{\pi}{2},
\end{equation}
then
\begin{eqnarray}
l^{a}&=&(0,k^{0},0,0),
\\
m^{a}&=&(0,0,k^{0},0),
\\
n^{a}&=&(k^{3},0,0,k^{0}).
\end{eqnarray}
Thus $(a\cdot n)$ is the unphysical polarization. From Eq.(\ref{polar}), 
the $r$-polarization $(a\cdot l)$ and the $\theta$-polarization 
$(a\cdot m)$ satisfy the equations,
\begin{eqnarray}
&&\left\{k^{2}-\frac{1}{2m^{2}}be^{2\hat{a}\phi}
\left[Al^{2}+B(m^{2}+n^{2})+Cq^{2}+Dr^{2}\right]\right.
\nonumber\\
&&\ -\frac{2}{m^{2}}ce^{2\hat{a}\phi}\left[A
+C\left(\frac{k^{3}}{k^{0}}\right)^{2}\right]l^{2}
\left.-\frac{8}{m^{4}}ze^{2\hat{a}\phi}
\left(\frac{Q}{4\pi r^{2}}\right)^{2}l^{2}
-\frac{4}{m^{4}}ye^{2\hat{a}\phi}
\left(\frac{Q}{4\pi r^{2}}\right)^{2}l^{2}\right\}(a\cdot l)=0,
\nonumber\\
&&
\end{eqnarray}
and
\begin{eqnarray}
&&\left\{k^{2}-\frac{1}{2m^{2}}be^{2\hat{a}\phi}
\left[Al^{2}+B(m^{2}+n^{2})+Cq^{2}+Dr^{2}\right]\right.
\nonumber\\
&&\ -\frac{2}{m^{2}}ce^{2\hat{a}\phi}\left[B
+D\left(\frac{k^{3}}{k^{0}}\right)^{2}\right]m^{2}
\left.-\frac{2}{m^{4}}ye^{2\hat{a}\phi}
\left(\frac{Q}{4\pi r^{2}}\right)^{2}l^{2}\right\}(a\cdot m)=0.
\end{eqnarray}
For the polarizations $(a\cdot l)$ or $(a\cdot m)$ to be
non-zero, we obtain the photon velocities for these two
polarizations:
\begin{eqnarray}
\left\vert\frac{k^{0}}{k^{3}}\right\vert_{r\  {\rm pol}}
&=&1-\frac{1}{4m^{2}}be^{2\hat{a}\phi}(A+B+C+D)-
\frac{1}{m^{2}}ce^{2\hat{a}\phi}(A+C)
\nonumber\\
&&\ -\frac{4}{m^{4}}ze^{2\hat{a}\phi}
\left(\frac{Q}{4\pi r^{2}}\right)^{2}
-\frac{2}{m^{4}}ye^{2\hat{a}\phi}
\left(\frac{Q}{4\pi r^{2}}\right)^{2},
\label{rpol}
\\
\left\vert\frac{k^{0}}{k^{3}}\right\vert_{\theta\  {\rm pol}}
&=&1-\frac{1}{4m^{2}}be^{2\hat{a}\phi}(A+B+C+D)-
\frac{1}{m^{2}}ce^{2\hat{a}\phi}(B+D)
\nonumber\\
&&\ -\frac{1}{m^{4}}ye^{2\hat{a}\phi}
\left(\frac{Q}{4\pi r^{2}}\right)^{2}.
\label{tpol}
\end{eqnarray}
Therefore, the light-cone condition is also modified for 
orbital photons, and the velocities of the photons for the
two polarizations are different, a phenomenon of gravitational
birefringence \cite{DH}.

To see how the light-cone condition is modified, we shall examine
more closely the RN case with $\hat{a}^{2}=0$ and
the stringy case with $\hat{a}^{2}=1$. 
For $\hat{a}^{2}=0$, 
\begin{eqnarray}
\lambda&=&\left(1-\frac{r_{+}}{r}\right)^{1/2}
\left(1-\frac{r_{-}}{r}\right)^{1/2},
\\
R&=&r,
\end{eqnarray}
with
\begin{eqnarray}
2GM&=&r_{+}+r_{-},
\\
\frac{GQ^{2}}{4\pi}&=&r_{+}r_{-}.
\end{eqnarray}
Then
\begin{eqnarray}
A+B+C+D&=&\frac{GQ^{2}}{2\pi r^{4}},
\\
A+C&=&-\frac{3GM}{r^{3}}+\frac{GQ^{2}}{\pi r^{4}},
\\
B+D&=&\frac{3GM}{r^{3}}-\frac{GQ^{2}}{2\pi r^{4}},
\end{eqnarray}
and
\begin{eqnarray}
\left\vert\frac{k^{0}}{k^{3}}\right\vert_{r\  {\rm pol}}
&=&1-\frac{b}{4m^{2}}\left(\frac{GQ^{2}}{2\pi r^{4}}\right)
-\frac{c}{m^{2}}\left(-\frac{3GM}{r^{3}}
+\frac{GQ^{2}}{\pi r^{4}}\right)
\nonumber\\
&&\ -\frac{4z}{m^{4}}\left(\frac{Q}{4\pi r^{2}}\right)^{2}
-\frac{2y}{m^{4}}\left(\frac{Q}{4\pi r^{2}}\right)^{2},
\label{rnrpol}
\\
\left\vert\frac{k^{0}}{k^{3}}\right\vert_{\theta\  {\rm pol}}
&=&1-\frac{b}{4m^{2}}\left(\frac{GQ^{2}}{2\pi r^{4}}\right)
-\frac{c}{m^{2}}\left(\frac{3GM}{r^{3}}
-\frac{GQ^{2}}{2\pi r^{4}}\right)
\nonumber\\
&&\ -\frac{1}{m^{4}}y
\left(\frac{Q}{4\pi r^{2}}\right)^{2},
\label{rntpol}
\end{eqnarray}
which are the same as the results in \cite{DS1}. For QED corrections,
Eqs.(\ref{rnrpol}) and (\ref{rntpol}) become 
\begin{eqnarray}
\left\vert\frac{k^{0}}{k^{3}}\right\vert_{r\  {\rm pol}}
&=&1+\frac{1}{30}\left(\frac{\alpha}{\pi}\right)
\left(\frac{1}{m^{2}}\right)\left(\frac{GM}{r^{3}}\right)
+\frac{1}{36}\left(\frac{\alpha}{\pi}\right)
\left(\frac{1}{m^{2}}\right)\left(\frac{GQ^{2}}{4\pi r^{4}}\right)
\nonumber\\
&&\ -\frac{8}{45}\left(\frac{\alpha^{2}}{m^{4}}\right)
\left(\frac{Q}{4\pi r^{2}}\right)^{2},
\\
\left\vert\frac{k^{0}}{k^{3}}\right\vert_{\theta\  {\rm pol}}
&=&1-\frac{1}{30}\left(\frac{\alpha}{\pi}\right)
\left(\frac{1}{m^{2}}\right)\left(\frac{GM}{r^{3}}\right)
+\frac{17}{180}\left(\frac{\alpha}{\pi}\right)
\left(\frac{1}{m^{2}}\right)\left(\frac{GQ^{2}}{4\pi r^{4}}\right)
\nonumber\\
&&\ -\frac{14}{45}\left(\frac{\alpha^{2}}{m^{4}}\right)
\left(\frac{Q}{4\pi r^{2}}\right)^{2}.
\end{eqnarray}
To compare the magnitudes of various terms, we define,
following Daniels and Shore \cite{DS1}, the accretion limit
charge
\begin{equation}
Q_{0}\equiv\sqrt{\frac{4\pi}{\alpha}}GMm.
\end{equation}
In terms of $Q_{0}$, 
\begin{eqnarray}
\left\vert\frac{k^{0}}{k^{3}}\right\vert_{r\ {\rm pol}}
&=&1+\frac{1}{240}\left(\frac{\alpha}{\pi}\right)
\frac{1}{(GMm)^{2}}\left(\frac{2GM}{r}\right)^{3}
\left[1+\frac{5}{12}\left(\frac{Gm^{2}}{\alpha}\right)
\left(\frac{Q}{Q_{0}}\right)^{2}\left(\frac{2GM}{r}\right)
\right.
\nonumber\\
&&\ \ \ \ \ \ \ \ \ \ \left.-\frac{2}{3}\left(\frac{Q}{Q_{0}}\right)^{2}
\left(\frac{2GM}{r}\right)\right],
\label{q01}
\\
\left\vert\frac{k^{0}}{k^{3}}\right\vert_{\theta\ {\rm pol}}
&=&1+\frac{1}{240}\left(\frac{\alpha}{\pi}\right)
\frac{1}{(GMm)^{2}}\left(\frac{2GM}{r}\right)^{3}
\left[-1+\frac{17}{12}\left(\frac{Gm^{2}}{\alpha}\right)
\left(\frac{Q}{Q_{0}}\right)^{2}\left(\frac{2GM}{r}\right)
\right.
\nonumber\\
&&\ \ \ \ \ \ \ \ \ \ \left.-\frac{7}{6}\left(\frac{Q}{Q_{0}}\right)^{2}
\left(\frac{2GM}{r}\right)\right].
\label{q02}
\end{eqnarray}
In the square brackets, the three terms represent, respectively,
the gravitational effect identical to the Schwarzschild case, 
the indirect effect of the charge due to its modification of the
gravitational field, and the contribution of the electromagnetic
field itself. The second term includes a factor
\begin{equation}
\frac{Gm^{2}}{\alpha}\simeq 10^{-43}
\end{equation}
so that it is much smaller than the third term. On
the other hand, for the gravitational contribution to be 
comparable to the electromagnetic one, one must require
\begin{equation}
Q\simeq Q_{0}
\end{equation}
for $r\simeq 2GM$. Then the first and the third terms are of the
same magnitude,
\begin{eqnarray}
\left\vert\frac{k^{0}}{k^{3}}\right\vert_{r\ {\rm pol}}
&=&1+\frac{1}{240}\left(\frac{\alpha}{\pi}\right)
\frac{1}{(GMm)^{2}}\left(\frac{2GM}{r}\right)^{3}
\left[1-\frac{2}{3}\left(\frac{Q}{Q_{0}}\right)^{2}
\left(\frac{2GM}{r}\right)\right],
\\
\left\vert\frac{k^{0}}{k^{3}}\right\vert_{\theta\ {\rm pol}}
&=&1-\frac{1}{240}\left(\frac{\alpha}{\pi}\right)
\frac{1}{(GMm)^{2}}\left(\frac{2GM}{r}\right)^{3}
\left[1+\frac{7}{6}\left(\frac{Q}{Q_{0}}\right)^{2}
\left(\frac{2GM}{r}\right)\right].
\end{eqnarray}
For $\theta$-polarization, the velocity is always smaller than
$c$. For $r$-polarization, the velocity could be larger than
$c$ if 
\begin{eqnarray}
&&1-\frac{2}{3}\left(\frac{Q}{Q_{0}}\right)^{2}\frac{2GM}{r}>0
\nonumber\\
&\Longrightarrow&\frac{r}{2GM}>\frac{2}{3}\left(\frac{Q}{Q_{0}}\right)^{2},
\end{eqnarray}
which is again the same conclusion as in \cite{DS1}.

For comparison, we note that the extremal value $Q_{ext}$ for a RN
black hole is given by
\begin{equation}
Q_{ext}=\sqrt{4\pi GM^{2}},
\end{equation}
and
\begin{equation}
\frac{Q_{0}}{Q_{ext}}=\sqrt{\frac{Gm^{2}}{\alpha}}.
\end{equation}
Hence for the gravitational effect to be of significance, the charge of the
black hole must be much smaller than the extremal value,
that is, 
\begin{equation}
Q\simeq Q_{0}\ll Q_{ext}.
\end{equation}
However, if $Q\simeq Q_{ext}\gg Q_{0}$, the electromagnetic
term dominates. Then Eqs.(\ref{q01}) and (\ref{q02}) become
\begin{eqnarray}
\left\vert\frac{k^{0}}{k^{3}}\right\vert_{r\ {\rm pol}}
&=&1-\frac{1}{360}\left(\frac{\alpha}{\pi}\right)
\frac{1}{(GMm)^{2}}\left(\frac{2GM}{r}\right)^{4}
\left(\frac{Q}{Q_{0}}\right)^{2},
\label{erpol}
\\
\left\vert\frac{k^{0}}{k^{3}}\right\vert_{\theta\ {\rm pol}}
&=&1-\frac{7}{1440}\left(\frac{\alpha}{\pi}\right)
\frac{1}{(GMm)^{2}}\left(\frac{2GM}{r}\right)^{4}
\left(\frac{Q}{Q_{0}}\right)^{2},
\label{etpol}
\end{eqnarray}
for $r\simeq 2GM$. The velocities for both polarizations are 
subluminal but different from each other, 
and this is just the phenomenon of electromagnetic
birefringence \cite{SLA}.

Next we turn to the stringy case, $\hat{a}^{2}=1$, which we take
as a typical example for non-zero $\hat{a}^{2}$. With 
$\hat{a}^{2}=1$,
\begin{eqnarray}
r_{+}&=&2GM,
\\
r_{-}&=&Q^{2}/4\pi M,
\end{eqnarray}
then
\begin{eqnarray}
e^{2\phi}(A+B+C+D)
&=&\frac{GQ^{2}}{2\pi r^{4}},
\label{atod}
\\
e^{2\phi}(A+C)
&=&\left(GM-\frac{GQ^{2}}{4\pi r}\right)^{-1}
\left(\frac{1}{4r^{3}}\right)
\left[-12(GM)^{2}+22(GM)\left(\frac{GQ^{2}}{4\pi r}\right)
\right.
\nonumber\\
&&\ \ \ \ \ \ \ \ \ \ \left.-12\left(\frac{GQ^{2}}{4\pi r}\right)^{2}
+\left(\frac{r}{GM}\right)\left(\frac{GQ^{2}}{4\pi r}\right)^{2}
\right],
\label{ac}
\\
e^{2\phi}(B+D)
&=&\left(GM-\frac{GQ^{2}}{4\pi r}\right)^{-1}
\left(\frac{1}{4r^{3}}\right)
\left[12(GM)^{2}-14(GM)\left(\frac{GQ^{2}}{4\pi r}\right)
\right.
\nonumber\\
&&\ \ \ \ \ \ \ \ \ \ \left.+4\left(\frac{GQ^{2}}{4\pi r}\right)^{2}
-\left(\frac{r}{GM}\right)\left(\frac{GQ^{2}}{4\pi r}\right)^{2}
\right],
\label{bd}
\\
e^{2\phi}
&=&\frac{1}{GM}\left(GM-\frac{GQ^{2}}{4\pi r}\right).
\label{phi}
\end{eqnarray}

First we assume that the charge
\begin{equation}
Q\simeq Q_{0}\ll Q_{ext}
\end{equation}
where the extremal condition $r_{+}=r_{-}$ now gives
\begin{equation}
Q_{ext}=\sqrt{8\pi GM^{2}}.
\end{equation}
Hence, $Q\ll Q_{ext}$ implies that
\begin{equation}
GM\gg\frac{GQ^{2}}{4\pi r}
\end{equation}
for $r\simeq 2GM$. In this approximation, we can expand the expressions
in Eqs.(\ref{ac}) and (\ref{bd}),
\begin{eqnarray}
e^{2\phi}(A+C)&=&-\frac{3GM}{r^{3}}
+\frac{5GQ^{2}}{8\pi r^{4}}+\cdots,
\\
e^{2\phi}(B+D)&=&\frac{3GM}{r^{3}}
-\frac{GQ^{2}}{8\pi r^{4}}+\cdots.
\end{eqnarray}
Keeping only leading terms in the expressions above, Eqs.(\ref{rpol})
and (\ref{tpol}) become
\begin{eqnarray}
\left\vert\frac{k^{0}}{k^{3}}\right\vert_{r\  {\rm pol}}
&=&{\rm RN\  result}+\frac{c}{m^{2}}
\left(\frac{3GQ^{2}}{8\pi r^{4}}\right)
+\frac{4z}{m^{4}}\left(\frac{GQ^{2}}{4\pi rGM}\right)
\left(\frac{Q}{4\pi r^{2}}\right)^{2}
\nonumber\\
&&\ \ \ \ \ +\frac{2y}{m^{4}}
\left(\frac{GQ^{2}}{4\pi rGM}\right)
\left(\frac{Q}{4\pi r^{2}}\right)^{2}
\\
\left\vert\frac{k^{0}}{k^{3}}\right\vert_{\theta\  {\rm pol}}
&=&{\rm RN\  result}-\frac{c}{m^{2}}
\left(\frac{3GQ^{2}}{8\pi r^{4}}\right)
+\frac{y}{m^{4}}\left(\frac{GQ^{2}}{4\pi rGM}\right)
\left(\frac{Q}{4\pi r^{2}}\right)^{2}
\end{eqnarray}
For QED corrections,
\begin{eqnarray}
\left\vert\frac{k^{0}}{k^{3}}\right\vert_{r\ {\rm pol}}
&=&{\rm RN\  result}+\frac{1}{240}\left(\frac{\alpha}{\pi}\right)
\frac{1}{(GMm)^{2}}\left(\frac{2GM}{r}\right)^{3}
\left[\frac{1}{4}\left(\frac{Gm^{2}}{\alpha}\right)
\left(\frac{Q}{Q_{0}}\right)^{2}\left(\frac{2GM}{r}\right)
\right.
\nonumber\\
&&\ \ \ \ \ \ \ \ \ \ \left.+\frac{1}{3}\left(\frac{Gm^{2}}{\alpha}\right)
\left(\frac{Q}{Q_{0}}\right)^{4}\left(\frac{2GM}{r}\right)^{2}\right],
\\
\left\vert\frac{k^{0}}{k^{3}}\right\vert_{\theta\ {\rm pol}}
&=&{\rm RN\  result}+\frac{1}{240}\left(\frac{\alpha}{\pi}\right)
\frac{1}{(GMm)^{2}}\left(\frac{2GM}{r}\right)^{3}
\left[-\frac{1}{4}\left(\frac{Gm^{2}}{\alpha}\right)
\left(\frac{Q}{Q_{0}}\right)^{2}\left(\frac{2GM}{r}\right)
\right.
\nonumber\\
&&\ \ \ \ \ \ \ \ \ \ \left.+\frac{7}{12}\left(\frac{Gm^{2}}{\alpha}\right)
\left(\frac{Q}{Q_{0}}\right)^{4}\left(\frac{2GM}{r}\right)^{2}\right].
\end{eqnarray}
The extra terms are subleading with respect to the RN results because
they are proportional to $Gm^{2}/\alpha\simeq 10^{-43}$. Therefore,
conclusions from the RN case will not be altered. 
The smallness of the dilatonic effects in this case can be understood
from the fact that for the dilaton black holes that we are considering,
the dilaton charge is given by \cite{GHS}
\begin{equation}
D=\frac{1}{4\pi}\oint_{S}d^{2}S^{\mu}\ \nabla_{\mu}\phi
=\frac{Q^{2}}{8\pi GM}.
\label{dcharge}
\end{equation}
Hence, for $Q\ll Q_{ext}=\sqrt{8\pi GM^{2}}$,
\begin{equation}
D\ll \frac{Q_{ext}^{2}}{8\pi GM}=M.
\label{dchargem}
\end{equation}
The dilatonic effect is therefore much smaller than the gravitational
effect. 

On the other hand, if $Q\simeq Q_{ext}$, or $D\simeq M$,
the dilatonic effect should be of importance. Thus we shall now 
look at this extremal case, with
\begin{equation}
Q=Q_{ext}=\sqrt{8\pi GM^{2}}.
\label{extreme}
\end{equation}
Remember that in the RN case, the electromagnetic effect dominates
over the gravitational one for $Q\simeq Q_{ext}\gg Q_{0}$ and
$r\simeq 2GM$, as in Eqs.(\ref{erpol}) and (\ref{etpol}).

With Eq.(\ref{extreme}),
Eqs.(\ref{atod}) to (\ref{phi}) simplify to
\begin{eqnarray}
e^{2\phi}(A+B+C+D)&=&\frac{1}{r^{2}}
\left(\frac{2GM}{r}\right)^{2},
\\
e^{2\phi}(A+C)&=&-\frac{3}{2r^{2}}
\left(\frac{2GM}{r}\right)\left(1-\frac{2GM}{r}\right),
\\
e^{2\phi}(B+D)&=&\frac{1}{2r^{2}}
\left(\frac{2GM}{r}\right)\left(3-\frac{2GM}{r}\right),
\\
e^{2\phi}&=&1-\frac{2GM}{r}.
\end{eqnarray}
Then,
\begin{eqnarray}
\left\vert\frac{k^{0}}{k^{3}}\right\vert_{r\ {\rm pol}}
&=&1-\frac{b}{4m^{2}}\left(\frac{1}{r^{2}}\right)
\left(\frac{2GM}{r}\right)^{2}
+\frac{c}{m^{2}}\left(\frac{3}{2r^{2}}\right)
\left(\frac{2GM}{r}\right)\left(1-\frac{2GM}{r}\right)
\nonumber\\
&&\ -\frac{2}{m^{4}}(2z+y)
\left(\frac{1}{8\pi Gr^{2}}\right)
\left(\frac{2GM}{r}\right)^{2}\left(1-\frac{2GM}{r}\right),
\\
\left\vert\frac{k^{0}}{k^{3}}\right\vert_{\theta\ {\rm pol}}
&=&1-\frac{b}{4m^{2}}\left(\frac{1}{r^{2}}\right)
\left(\frac{2GM}{r}\right)^{2}
-\frac{c}{m^{2}}\left(\frac{1}{2r^{2}}\right)
\left(\frac{2GM}{r}\right)\left(3-\frac{2GM}{r}\right)
\nonumber\\
&&\ -\frac{y}{m^{4}}
\left(\frac{1}{8\pi Gr^{2}}\right)
\left(\frac{2GM}{r}\right)^{2}\left(1-\frac{2GM}{r}\right).
\end{eqnarray}
For QED corrections,
\begin{eqnarray}
\left\vert\frac{k^{0}}{k^{3}}\right\vert_{r\ {\rm pol}} 
&=&1+\left(\frac{\alpha}{\pi}\right)
\left(\frac{1}{m^{2}r^{2}}\right)
\left(\frac{2GM}{r}\right)
\left[\frac{13}{360}\left(\frac{2GM}{r}\right)
+\frac{1}{60}\left(1-\frac{2GM}{r}\right)
\right.
\nonumber\\
&&\ \left.-\frac{1}{45}\left(\frac{\alpha}{Gm^{2}}\right)
\left(\frac{2GM}{r}\right)\left(1-\frac{2GM}{r}\right)\right],
\label{drpol}
\\
\left\vert\frac{k^{0}}{k^{3}}\right\vert_{\theta\ {\rm pol}} 
&=&1+\left(\frac{\alpha}{\pi}\right)
\left(\frac{1}{m^{2}r^{2}}\right)
\left(\frac{2GM}{r}\right)
\left[\frac{13}{360}\left(\frac{2GM}{r}\right)
-\frac{1}{180}\left(3-\frac{2GM}{r}\right)
\right.
\nonumber\\
&&\ \left.-\frac{7}{180}\left(\frac{\alpha}{Gm^{2}}\right)
\left(\frac{2GM}{r}\right)\left(1-\frac{2GM}{r}\right)\right].
\label{dtpol}
\end{eqnarray}
In most cases the last terms in the square brackets in 
Eqs.(\ref{drpol}) and (\ref{dtpol}), 
which come from the electromagnetic part, still dominate 
because of the factor of $\alpha/Gm^{2}$. However, when 
$r=2GM$, these terms vanish, and
\begin{eqnarray}
\left\vert\frac{k^{0}}{k^{3}}\right\vert_{r\ {\rm pol}}
&=&1+\frac{13}{1440}\left(\frac{\alpha}{\pi}\right)
\frac{1}{(GMm)^{2}},
\\
\left\vert\frac{k^{0}}{k^{3}}\right\vert_{\theta\ {\rm pol}}
&=&1+\frac{1}{160}\left(\frac{\alpha}{\pi}\right)
\frac{1}{(GMm)^{2}}.
\end{eqnarray}
This indicates that the dilatonic effect in fact becomes the most
dominant one at the event horizon, and the velocities of the photons
there in both polarizations are superluminal. There 
is a caveat here though. For $\hat{a}^{2}>0$, the inner horizon
is a singular surface, and in the extremal case the inner and
event horizons merge. As one approaches the event horizon, which
is now a singular surface, the curvature becomes very large. This
would contradict our assumption that $L\gg \lambda$. 
Thus the conclusion above can at best be taken only as an indication
that the dilatonic effect would become more and more important for
determining the light-cone condition of orbital photons as 
one gets nearer to the event horizon
in the extremal or near-extremal cases.

\section{Conclusions and discussions}
We have investigated the phenomenon of ``faster than light"
photons, which occurs due to the matter quantum corrections to
the photon propagator, in a family of dilaton black holes
parametrized by $\hat{a}^{2}$. For $\hat{a}^{2}=0,1$ and 3, we 
have the usual RN, the stringy \cite{GHS} 
and the Kaluza-Klein black holes \cite{GM},
respectively. Dilaton black holes are particular examples of 
black holes with scalar hairs, and their singularity structures
are also quite different. It is therefore interesting to see if 
peculiar behaviors are present when this phenomenon is considered
in this case.

Indeed we find that for radially directed photons, the light-cone
condition is modified despite the fact that the spacetimes are 
spherically symmetric. This result is different from the cases of 
Schwarzschild and RN black holes 
in which the light-cone condition
for radial photons are unchanged. In this dilatonic case, 
the light-cone condition of the radial photons also
satisfy the ``horizon theorem" and the ``polarization sum rule" 
of Shore \cite{GMS}.

For orbital photons, we concentrate on the stringy case, 
$\hat{a}^{2}=1$, as a typical example for non-zero $\hat{a}^{2}$. 
With the charge $Q\simeq Q_{0}\ll Q_{ext}$, where $Q_{0}$ is the
accretion limit charge and $Q_{ext}$ the extremal charge,
the pure gravitational effect (Schwarzschild part) is comparable
to the electromagnetic one. The dilaton charge is much small than
the mass, Eq.(\ref{dchargem}), so the dilatonic effect is negligible. 
On the other hand, if $Q\simeq Q_{ext}$, the electromagnetic effect 
will still dominate over the gravitational one, as in the RN case, 
so that the velocities of the photons in both polarizations are
subluminal.
The only exception is that 
as one approaches the event horizon, the dilatonic effect becomes
more and more important, even more so than the electromagnetic one,
and the velocities of the photons change to being superluminal.
This indicates that the dilatonic effect is crucial in determining
this ``faster than light" phenomenon when the photon is near the
event horizon in the extremal or near-extremal cases.

In arriving at the above results, we have taken the approximations
\begin{equation}
L\gg\lambda\gg\lambda_{c}
\end{equation}
where $L$ represents the scale of the typical variation of the 
background fields, $\lambda$ the photon wavelength and $\lambda_{c}$
the Compton wavelength. For $\hat{a}^{2}>0$, the inner horizon becomes
singular. In addition, the inner and the event horizons merge in the
extremal case. When the orbital photon is very near to the event
horizon as mentioned above, the curvature becomes so large that our
approximation may no longer be valid. Hence, we have taken this result
for the orbital photons only as an indication on the growing importance
of the dilatonic effect as one gets near to the event horizon.

It is therefore of interest to extend our investigation beyond the 
assumed range above if, for example, one wants to consider the extremal
black holes more carefully. 
To relax the condition $\lambda\gg\lambda_{c}$ requires
the summation of terms like $\sum_{i}(R/m^{2})(\nabla^{i}/m^{i})F^{2}$
in the effective action. As discussed in Ref. 3, this may be achieved
by the new summation technique of Barvinsky and Vilkovisky \cite{BV}. 
Whereas to go beyond the other constraint $L\gg\lambda$, we need to
push the formalism to the strong field regime and 
one must then use non-perturbative approximations. 
For instance, one may try to consider
the light-cone condition of a massless fermion in the Gross-Neveu model
\cite{GN} in curved space by the $1/N$ approximation  to extract
non-perturbative effects.

In Section V, we have concentrated on the dilaton black holes with
$\hat{a}^{2}=1$ because this corresponds to the stringy case, which
is most interesting to us. Moreover, this case should be typical enough 
that similar results are expected for other non-zero values of
$\hat{a}^{2}$. On the other hand, the low-energy effective action
of string theories has introduced a whole new set of black holes \cite{RRH}.
The ones that we have considered here are the simplest. We can therefore
extend our consideration to, for instance, 
dyonic dilaton black holes \cite{GM,CLH},
black holes with axions \cite{MJB}, 
rotating dilaton black holes \cite{HH1} and even black
holes with non-trivial dilaton potentials \cite{GH,HH2}. 

\acknowledgments

We would like to thank C.-C. Chen and R.-R. Hsu for useful discussions.
This work was supported by the National Science Council of the Republic
of China under contract number NSC 86-2112-M032-001.


\begin{references}

\bibitem[1]{DH}I. T. Drummond and S. J. Hathrell, Phys. Rev. {\bf D22},
	       343 (1980).
\bibitem[2]{SLA}S. L. Adler, Ann. Phys. (N.Y.) {\bf 67}, 599 (1971).
\bibitem[3]{DS1}R. D. Daniels and G. M. Shore, Nucl. Phys. {\bf B425},
		634 (1994).
\bibitem[4]{DS2}R. D. Daniels and G. M. Shore, Phys. Lett. {\bf B367},
		75 (1996).
\bibitem[5]{GMS}G. M. Shore, Nucl. Phys. {\bf B460}, 379 (1996).                
\bibitem[6]{GM}G. Gibbons and K. Maeda, Nucl. Phys. {\bf B298},
	       741 (1988).
\bibitem[7]{GHS}D. Garfinkle, G. Horowitz, and A. Strominger, Phys. Rev.
		{\bf D43}, 3140 (1991); {\it ibid}, {\bf D45}, 3888 (1992).
\bibitem[8]{MTW}C. W. Misner, K. S. Thorne, and J. A. Wheeler, 
		{\it Gravitation} (Freeman, New York, 1973).
\bibitem[9]{BV}A. O. Barvinsky and G. A. Vilkovisky, Nucl. Phys.  
	       {\bf B282}, 163 (1987); {\it ibid}, {\bf B333}, 471 (1990);
	       {\it ibid}, {\bf B333}, 512 (1990).
\bibitem[10]{GN}D. Gross and A. Neveu, Phys. Rev. {\bf D10}, 3235 (1974).
\bibitem[11]{RRH}R.-R. Hsu, Chinese J. Phys. {\bf 30}, 569 (1992).
\bibitem[12]{CLH}G.-J. Cheng, W.-F. Lin, and R.-R. Hsu, J. Math. Phys.
		 {\bf 35}, 4839 (1994).
\bibitem[13]{MJB}M. J. Bowick {\it et al}, Phys. Rev. Lett. {\bf 61},
		 2823 (1988).
\bibitem[14]{HH1}J. H. Horne and G. T. Horowitz, Phys. Rev. {\bf D46},
		 1340 (1992).
\bibitem[15]{GH}R. Gregory and J. A. Harvey, Phys. Rev. {\bf D47},
		2411 (1993).
\bibitem[16]{HH2}J. H. Horne and G. T. Horowitz, Nucl. Phys. {\bf B399},
		 169 (1993).

\end{references}
\end{document}